\documentclass[useAMS,usenatbib]{mn2e}
\usepackage{amsmath}
\usepackage{graphicx}

\numberwithin{equation}{section}

\def\msol{{\rm ~M_\odot}}
\def\kms{{\rm km~s^{-1}}}

\def\Ro{R_{{o}}}

\def\Msol{M_{\odot}}

\title{Incorporating Streams into Milky Way Models}
\author[N. Deg \& L.Widrow]
  {{Nathan Deg$^{1}$\thanks{E-mail:ndeg@astro.queensu.ca},
  Lawrence Widrow$^{1}$}\\
  $^1$Department of Physics, Engineering Physics, and Astronomy,Queen's University, 
\\Kingston, ON, K7L 3N6, Canada}

\begin{document}

\maketitle

\begin{abstract}

We develop a framework for modelling the Milky Way using stellar 
streams and a wide range of photometric and kinematic observations.
Through the use of mock data we demonstrate that a 
standard suite of Galactic observations leads to degeneracies 
in the inferred halo parameters.  We then incorporate a 
GD-1-like stream into this suite using the 
orbit-fitting technique and show that the streams reduces
the uncertainties in these parameters provided all observations are
fit simultaneously.  We also explore 
how the assumption of a disk-halo alignment can lead 
unphysical models.  Our results may explain 
why some studies based on the Sagittarius stream find that the 
halo's intermediate axis is parallel to the disk spin axis even 
though such a configuration is highly unstable.
Finally we show that both longer streams and multiple 
streams lead to improvements in our ability to infer the shape 
of our dark halo.
\end{abstract}


\section{Introduction}

In principle, a detailed model of the Milky Way's (MW) gravitational potential can 
be used to infer the shape and structure of the Galaxy's 
dark matter (DM) halo.  The implications of such a determination range from
dark matter detection to cosmological structure formation to galaxy
evolution in the present-day Universe.

In any attempt to infer Galactic structure from data it is necessary to specify 
the space of Galactic models under consideration, $M$, the data, $D$, and the means 
by which one compares the two.  The latter is often expressed in 
terms of a likelihood function, $p(D|M)$.  In Bayesian statistics, one also specifies a 
prior probability on the model and then inverts the likelihood 
function to obtain the probability of the model given the data, $p(M|D)$.

Galaxy models are fraught with degeneracies, particularly when the set of observational
constraints are limited.  For example, models of external galaxies that are constrained by the 
rotation curve and surface brightness profile are plagued 
by the disk-halo degeneracy.  These data are found in the plane of the 
disk making them insensitive to the halo shape and structure, thereby
allowing one to trade off the disk and 
halo when fitting the predicted circular speed (see \citet{Courteau2013} and references therein).
The disk-halo degeneracy leads to uncertainties in the local DM density and the stability 
of the disk against the formation of a bar or spiral structure.

In principle model degeneracies can be broken by combining different types of observations
that sample different regions of the Galaxy.  
Stellar streams are a particularly promising and relatively new class of Galactic observations.
These roughly one-dimensional stellar features are presumably formed
when stars are stripped from dwarf galaxies by the tidal field of the host.  
They are located throughout the halo at a variety of radii which allows them 
to probe the shape and structure of the halo itself and break degeneracies like the 
disk-halo degeneracy.
The Sagittarius (Sgr) stream is perhaps the most prominent example of a stellar stream and
has been mapped over $\sim300^\circ$ across the sky.  Soon after its discovery
\citep{Ibata1997} astronomers began modelling the MW using the stream in
an effort to infer the Galactic potential. Some 
early examples include \citet{Ibata2001}, \citet{Helmi2004B},
\citet{Martinez2004}, \citet*{Johnston2005}, and \citet{Fellhauer2006}.

Recently
\citet{Law2009} utilized the Sgr stream to model the MW.  Their Galactic model consists
of a Hernquist bulge, a Miyamota-Nagai disk, and a triaxial logarithmic.  
The halo was oriented so that 
one of its symmetry axes coincides with the spin axis of the disk and 
is characterized by five parameters: the scale length, the scale velocity, the
axis ratios of the halo potential, and the orientation angle of the disk-plane symmetry axes.  Of these
parameters, \citet{Law2009} only explored the axis ratios and orientation.  The 
halo scale length, as well as the bulge and disk parameters, are fixed to 
preferred values based on previous work \citep{Law2005}.  The halo scale velocity 
was adjusted so that the circular speed remains constant.  The Sgr stream data used in their 
fit consisted of the M giant survey
from \citet{Majewski2003} and Sloan Digital Sky Survey (SDSS) observations from \citet{Belokurov2006}.  
The stream itself was modeled under the assumption that it traces the orbit of a single particle 
in a fixed potential (the orbit-fitting 
technique).  Through a grid search of their three free parameters 
they found that a favored model with isopotential axis ratios of $1.5:1.25:1.$

\citet{Law2010} refined the work of \citet{Law2009} by using N-body methods to model
formation of the stream.
They found that the dark halo had \textit{isopotential} axis ratios of $1.28:1.26:1$ 
with the intermediate axis perpendicular to 
the Galactic disk and the short axis roughly along the Sun-Galactic Center line.
\citet{DW13}, using a different Galactic model, performed a Bayesian 
analysis of the Sgr stream and a suite of other observational constraints 
and found the halo to have isodensity axis ratios of $3.3\pm0.7:2.7\pm0.4:1$
with almost the same orientation.
However, \citet{Ibata2013}  used the 
streak-line method \citep{Varghese2011,Kupper2012} for the stream model
and found that they could fit the stream observations using a Galactic model with 
a spherical, non-parametric halo.
Additionally, \citet{VeraCiro2013} used a similar model to \citet{Law2009} but allowed the 
halo shape to change with radius.  They found results similar to \citet{Law2010}, but, 
when they included the effect of the Large Magellanic Cloud on the stream, the inferred halo shape 
was only mildly triaxial.

The Sgr stream is particularly difficult to model.
It is dynamically hot and has a complicated structure.  \citet{Belokurov2014} recently 
showed that the trailing arm has a different apocenter than the 
leading arm.  \citet{Penarrubia2010} has shown that the stream structure depends on 
the progenitor's structure as well as the MW.  Moreover, 
\citet{Zhoa2004} has shown that the orbit of Sgr is unlikely unless it is 
particularly massive or fell in with a group.  As such, its orbital 
properties are likely quite complex and difficult to model.  Finally, 
the stream shows an apparent bifurcation in both the leading 
and trailing arms \citet{Fellhauer2006}.  It is unclear whether this bifurcation is 
due to the formation of the stream, from some secondary companion, or whether 
it is a completely different stream altogether \citep{Koposov2012}. 
No study has yet been able to fit all these observations.

A number of studies have used other observed streams to 
model the MW. \citet{Willet2009} found an orbit fit for the GD-1 stream (discovered by \citet{Grillmair2006}) but 
were unable to derive a constraint on the halo shape.  However, \citet{Koposov2010} obtained 
6-D phase space information on the stream.  Using the orbit fitting technique 
with this data, they found the total potential to be oblate 
with the short axis perpendicular to the disk.  They note that this flattening may be 
mostly due to the disk.  When they include the disk mass uncertainties they were unable to find 
robust constraints on the halo flattening.  \citet{Newberg2010} modeled the Orphan stream 
(independently discovered by \citet{Grillmair2006a} and \citet{Belokurov2007}) 
using the orbit-fitting technique, but found that 
a longer stream is needed to distinguish between a variety of different MW models as 
many of their model orbits diverged significantly past the edge of the data.
\citet{Lux2012} fit orbits for the NGC 5466 stream (discovered by
\citet{Belokurov2006b,Grillmair2006c}) in a variety of Galactic models.  They found that only 
orbits in oblate or triaxial potentials produce deviations in the angular position that 
mark turning points in the orbit.  There are hints that such kinks are present in 
the stream data, suggesting a non-spherical halo.  
\citet{Grillmair2006d} attempted to find orbits for the Palomar 5 (Pal 5) stream 
(discovered by \citet{Odenkirchen2001})and found 
that a spherical halo fit the data reasonably well.  
From this sample of studies, it is clear that MW models generated using stellar streams 
vary greatly.  These variations are due to differences in three 
key elements of the analysis: the space of models 
examined, the data used as constraints, and the method for modeling the stream.

Today there are roughly a dozen known Galactic streams (see \citet{Sanders2013a} for a 
summary).  
The launch of Gaia \citep{Perryman2001} will surely lead to the discovery of 
new streams and improvements in the observational constraints from known streams.
With Gaia in mind, we set out to
carefully examine how streams can be combined with other Galactic observations to 
break some of the existing degeneracies in the halo structure and shape parameters.
We utilize mock data for this examination and model streams using the orbit-fitting 
technique.  This allows us to sidestep the issue of 
how best to model streams and focus on issues relating to the space of Galactic 
models, the data used to constrain those models, and the construction the likelihood function.

This study is complementary to the recent work of \citet{Lux2013}.  They 
explore the biases that arise in the inferred halo shape due to the use of the orbit-fitting technique 
on realistic streams using mock data.  Both our work and \citet{Lux2013} use 
Markov Chain Monte Carlo (MCMC) techniques to explore a parameter space, but 
we utilize non-stream constraints and explore a substantially larger parameter space.

We begin in Section 2 by describing a three-component MW model.  In Section 3 
we present a suite of Galactic observations that constrain the model.  We also describe our procedure for generating 
mock data for a MW-like galaxy.  In Section 4 we examine some of 
the degeneracies that are found in the model parameters when we fit only the 
suite of standard constraints, as well as where those constraints arise.
In Section 5 we present the stream observations, stream fitting algorithm, and 
mock data.  In Section 6
we show that streams observations must be combined all other constraints 
simultaneously to get the best constraints possible for the halo shape.  We then 
demonstrate how the potentially erroneous assumption of a disk-halo alignment, 
can affect the inferred parameters.  We finish 
Section 6 by examining the halo shape and orientation when a GD-1-like 
stream is combined with the full suite of constraints and the full model parameter 
space is explored.  In Section 7 we show how the use of either a longer stream or 
multiple streams provide leads to improvements in our ability to infer the shape of the 
dark halo.

\section{Galactic Model}
We model the Galactic density as a S\'{e}rsic bulge, an exponential disk, and a triaxial 
Einasto halo \citet{Einasto1965}.  An alternative and popular approach is to model the Galactic 
potential in terms of analytic functions (some examples include \citet{Law2010},
\citet{VeraCiro2013} and \citet{Newberg2010}).
It is computationally simpler to calculate forces from an analytic potential than it is to 
obtain the force by integrating the density.  However, cosmological simulations 
suggest that it is the halo density rather than the potential that is triaxial
\citep*{Frenk1988,Franx1991,Warren1992,Jing2002,Allgood2006}.  Indeed a triaxial 
potential can lead to peanut shaped densities, and, in extreme cases, negative 
values for the density \citep{BT2008}.  For this reason we model the Galaxy using
density functions.

The bulge density that generates a S\'{e}rsic profile in projection is
 \citep{Prugniel1997,Terzic2005} 
\begin{equation}\label{eq:Sersic}
\rho_{b}(r) = \frac{\sigma_{b}^{2}C(n)}{4\pi G R_{e}^{2}}
\left(\frac{r}{R_{e}}
\right)^{-p}e^{-b(r/Re)^{1/n}}~,
\end{equation}
where $n$ is the S\'{e}rsic index, $p = 1 - 0.6097/n + 0.05563/n^2$, 
$C(n)=nb^{n(p-2)}\Gamma(n(2-p))$, and
$b=b(n)$ is chosen so that the radius $R_e$ encloses half the
total projected mass. An analytic expression for the associated potential can 
be found in \citet{Terzic2005}

The disk density is
\begin{equation}\label{eq:diskdensity}
\rho_d\left (R,\,z\right ) =~\frac{M_d}{4\pi R_d^2z_d}   
\, e^{-R/R_d}\, {\rm sech}^2\left (z/z_d\right )
\end{equation} 
where $M_d$, $R_d$, and $z_d$, are the disk mass, radial scale length,
and vertical scale height.
The Einasto halo density \citep{Einasto1965,Merritt2005} is
\begin{equation}
 \rho_{h}(r_{t})=\rho_{0}e^{-\frac{2}{\alpha}((r_{t}/r_{h})^{\alpha}-1)}~,
\end{equation}
where, $\rho_{0}$, $r_{h}$, and $\alpha$ are the halo scale density, scale radius, 
and a parameter to control the logarithmic slope respectively.  
The triaxial radius, $r_{t}$, in component notation, is
\begin{equation}
 t_{t,i}=R_{i,j}\Lambda_{j,k}r_{k},
\end{equation}
where $\mathbf{\Lambda}$ is a diagonal matrix with elements $(1,A,B)$ where $A\ge1\ge B$ and 
$\mathbf{R}$ is an Euler rotational matrix made of consecutive rotations about the $Z$, $Y$, and $X$ axes.  
This set up means that
$A$ is the major/intermediate axis ratio, $B$ is the minor/intermediate axis ratio, and the 
first two Euler angles give the angular position of the intermediate axis.  In our analysis 
we simply use logarithmic priors for $A$ and $B$.

The potential for the halo is found using the homeoid theorem (see
\citet{BT2008}).  When the coordinate axes are aligned 
with the symmetry axes, the halo potential is
\begin{equation}\label{Eq:homeoid}
\Phi(\mathbf{x})=-\pi G \frac{a_{2}a_{3}}{a_{1}}\int_{0}^{\infty}
d\tau \frac{\psi(\infty)-\psi(m)}
{\sqrt{(\tau+a_{1}^{2})(\tau+a_{2}^{2})(\tau+a_{3}^{2})}}~
\end{equation}
where $a_{i}$ are the axis ratios, 
\begin{equation}\label{Eq:M}
m^2=a_{1}^{2}\sum_{i=1}^{3}\frac{x_{i}^{2}}{a_{i}^{2}+\tau}
\end{equation}
is similar to the square of the ellipsoidal radius, and
\begin{equation}
\psi(m)=\int_{0}^{m^{2}}dm^{2}\rho(m^{2})~
\end{equation}
is an auxiliary function.  From the definition of the halo shape $a_{1}=1$, $a_{2}=A$,
and $a_{3}=B$.

The disk potential is found using the technique of \citet{Kuijken1995}.  
An analytic 'fake' disk density-potential pair,
$\left (\rho_{fd},\,\Phi_{fd}\right )$ is constructed so that
$\rho_{d}=\rho_{fd}+\rho_{r}$ and $\Phi_d = \Phi_{fd} + \Phi_r$ where
$\left (\rho_r,\,\Phi_r\right )$ is the density-potential pair of the
residual.  The fake disk is designed to account for the high-order
moments that arise due to the thinness of the disk.  The Poisson equation is solved
via spherical harmonics up to $l=2$ for $\Phi_{r}$ in an iterative scheme and summed with the 
analytic fake disk potential.

\section{The Likelihood Function, Observational Constraints, and Mock Data}

In this section we present the likelihood function that is used to evaluate 
a particular model.  We then discuss the various non-stream observations used for 
constraining the Galactic model as well as the generation of mock data.

\subsection{The Likelihood Function}

Bayesian statistics provides a method for calculating the 
probability of a particular model given some data.
From Bayes theorem, the model's probability, called the posterior is
\begin{equation}
 p(M|D,I)=\frac{p(M|I)p(D|M)}{p(D|I)}~,
\end{equation}
where $I$ represents prior information, $p(M|I)$ is the prior
probability on the model $M$, and 
$p(D|M)$ is the likelihood of the data given the model.  The term $p(D|I)$, often referred to as the
evidence, is essentially a normalization factor and does not enter
our calculations.

The prior probability is simply the product of the prior probabilities for each 
free parameter.  Similarly, the likelihood is the product of the likelihood for 
each individual data point, regardless of the type of data, i.e. angular position, radial 
velocity, etc.  We assume that all errors are Gaussian 
so that the likelihood of any given data point is simply
\begin{equation}
 p(D_{i,j}|M)=\frac{1}{\sqrt{2\pi\sigma_{i,j}^{2}}}e^\frac{M_{i,j}-D_{i,j}}{2\sigma_{i,j}^{2}}~,
\end{equation}
where the $D_{i,j}$ is the $i$'th data point of the $j$'th type of observational 
constraint, $\sigma_{i,j}$ is the corresponding error, and $M_{i,j}$ is the 
model prediction.  

In order to fully map out the posterior distribution function (PDF) of the large parameter space
we utilize the EMCEE algorithm from \citet{Foreman-Mackey2012}.  This Markov Chain Monte Carlo (MCMC)
scheme is based on the 
Stretch-Move affine invariant algorithm found in \citet{Goodman2010} and scales particularly well 
with the number of parameters.  The chain is comprised of an ensamble of 
'walkers' that explore the parameter space.  The proposal for a given walker is
generated in two steps.  First, a second walker is randomly chosen.  Then a 
line in parameter space is 
drawn connecting the current state of the two walkers.  Finally a random number is 
selected from a square-root distribution to determine the length along 
the line one 'stretches' to select the proposal.

We modify the EMCEE algorithm in two ways.  Firstly we 
account for the fact that the Euler angles are essentially 
the product of a sphere with a circle.  We describe the modification
needed to deal with this non-trivial topology in the Appendix.  Secondly,
we include simulated annealing for stream constraints \citep{Gregory2005}.  We found that 
the stream constraints often contained a number of false minima and the chain became stuck.
Simulated annealing slowly cools the stream likelihood, which allows the chain to avoid being trapped 
by these false minima.  If we define the likelihood of some type of observation as
$\mathcal{L}_{j}=\sum_{i}\ln(p(D_{i,j}|M))$ then, with annealing, the likelihood is
\begin{equation}
 \mathcal{L}=\sum_{j}^{suite}\mathcal{L}_{j}+\beta(t)\sum_{j}^{streams}\mathcal{L}_{j}~,
\end{equation}
where $\beta(t)$ is the annealing temperature.  The total likelihood is 
$p(D|M)=e^\mathcal{L}$.  We slowly increase $\beta$ from 
$10^{-4}$ to $1$ in logarithmic intervals over 1000 MCMC steps.  The simulated 
annealing closely resembles the parallel tempering found in \citet{Varghese2011}.

\subsection{Observational Constraints}

In this work we use the suite of non-stream constraints found in \citet{DW13}, which 
followed the work of \citet{Widrow2008} and \citet{Dehnen1998}, and 
comprises observations of the Oort constants, 
the local circular speed, the local surface density and vertical force, the inner and 
outer rotation curves, the bulge surface brightness and line-of-sight velocity dispersion, 
and the total mass within 100 kpc.

Briefly, \citet{Reid2009} used Very Long Baseline radio
observations to determine trigonometric parallaxes and proper motions
for masers throughout the Milky Way's disk.  \citet{Bovy2009} reanalyzed these in 
conjunction with the proper motion of Sgr A$^{*}$ to find a circular speed 
of  $v_c\left(\Ro\right ) = 244\pm\,13\,\kms$.  
For the Oort constants, we use 
$A=14.8\pm0.8~\kms \textrm{kpc}^{-1}$ and $B=-12.4\pm0.6~\kms
\textrm{kpc}^{-1}$ from \citet{Feast1997}.  We use the disk surface
density measurement of $\Sigma_{d}=49\pm9~\msol~\rm{pc}^{-3}$ from
\citet{Flynn1994} and the vertical force measurement of
$|K_{z}(1.1~\textrm{kpc})| = \left ({2\pi G}\right )71\pm 6 ~\msol
\textrm{pc}^{-2}~,$ from \citet{Kuijken1991}, which are in
good agreement with similar studies (see, for example,
\citet{Holmberg2004}).  We will point out that the observations 
of the surface density are made with a 'rotation curve' prior
that is equivalent to assuming a spherical halo.  However, this 
study uses mock data self-consistently generated from 
a non-spherical model so any issues related to this prior 
are avoided.

For the bulge LOSVD we use the compilation of observations found in \citet{Tremaine2002} with 
the restriction that $r\ge4 ~\rm{pc}$ to avoid complications from the central black 
hole.  Additionally, we adjust the dispersion upwards by a factor of 1.07 to account 
for the non-sphericity of the bulge. For the surface brightness
profile we use the infrared COBE-DIRBE observations
\citep*{Binney1997}.

As is typical, we break the rotation curve constraints into an inner and outer portion.
For the inner rotation we fit the peak velocities along particular lines-of sight.  If the 
potential is axisymmetric the peak velocity is related to the circular 
speed by
$v_{\rm term} = v_c\left (R\right ) - v_c\left (\Ro\right )\sin{l}$.  This approximation still 
holds for a triaxial model as the axisymmetric disk and bulge components dominate the 
potential of the inner disk.  The 
data used is from \citet{Malhotra1995} with the restriction that $\sin l\ge 0.3$ 
so as to avoid distortions due to the bar.

The outer rotation curve is slightly more complicated.  The line-of-sight
velocity, $v_{lsr}$, is related to the circular speed in the expression
\begin{equation}
W(R)=\frac{\Ro}{R}\,v_{c}(R)-v_{c}(\Ro)=\frac{v_{lsr}}{\cos b \cos
  l}~.
\end{equation}
The data used consists of $v_{lsr}$ observations of HII regions and reflection nebulae from
\citet{Brand1993} and Carbon stars from \citet{Demers2007}.  In order
to avoid complications due to non-circular motions the data is restricted to
 $l\le 155^\circ$ or $l\ge 205^\circ$, $d\ge 1 $ kpc,
and $W\le 0$.  Additional 'noise' parameters are added to $d$ and $v_{lsr}$ to account for 
any unknown systematics.  Therefore the error for the $j$'th
data point is $\Sigma_{u,d}^{2}=\sigma^{2}_{u,d}+\epsilon^{2}_{u,d}$ where 
$\sigma$ is the observed error, $\epsilon$ is the noise parameter, and the 
$u$ or $d$ subscripts are for the distance and velocity respectively.  Since 
the systematics may be different between the two data sets we use two sets of 
noise parameters.  These calculated errors are then propagated through to errors
in $W$.

\subsection{Mock Data}

The mock data is generated from 
a model that is designed to resemble the MW.
To get a generative MW-like model we start by selecting a reasonable halo shape.
Figure 1 of \citet{Allgood2006}
shows that halos with masses $\simeq 10^{12} \Msol$
typically have a minor to major axis ratio of $\approx 0.60 \pm 0.02$.  Therefore we set
the generative model's axis ratios to $A=1.25$ and $B=0.75$.  We also set the orientation so that the 
disk and halo are misaligned.  The generative Euler angles are $(45^{\circ},30^{\circ},10^{\circ})$
which sets the halo's intermediate axis to be $45^{\circ}$ from the Sun-Galactic Center line 
with a  $30^{\circ}$ inclination.

The rest of the parameters for the generative model are found by analyzing the 
suite of non-stream observations.  The halo shape is fixed to that of the generative model
and we run the EMCEE algorithm to sample the remaining parameter space.
The model with the largest posterior is then selected to be the 
generative model.  After rounding 
the results, the generative model's parameters are set to $n=1$, $\sigma_{b}=225~\kms$, $R_{e}=0.8~\textrm{kpc}$,
$M_{d}=3.7\times10^{10}~\msol$, $R_{d}=2.7~\textrm{kpc}$, $z_{d}=0.25~\textrm{kpc}$,
$M/L_{b}=0.55$, $M/L_{d}=0.8$, $\rho_{0}=2.325\times10^{-3}~\msol~pc^{-3}$, $r_{h}=15~\textrm{kpc}$, and $\alpha=0.2$.
In addition to Galactic model's parameters, we also fit the 
Sun's location and peculiar velocity and find the peak at $R_{0}=8~\textrm{kpc}$, 
$(u_{\odot},v_{\odot},w_{\odot})=(-10~\kms,-5.45~\kms, 7.0~\kms)$.

The procedure for generating mock data is similar to the procedure for 
calculating the likelihood.  Just like the likelihood, we begin by 
calculating the model predictions for each observation.  Then we add
an error term from the Gaussian corresponding to the observed error.  This 
gives a mock data set with the same dispersion as the actual data.

\section{Standard Suite Exploration}

In this section we explore degeneracies in the halo model that arise in the 
absence of stream constraints.  We begin by analyzing
the mock data with the halo shape fixed to that of the generative model. 
The left panel of Figure 1 shows the resulting PDF for the halo scale length and density.
There is a clear degeneracy in these two parameters with the shape of the 
PDF being primarily determined by the local vertical force constraint.  As shown in the middle and right panels of 
Figure 1, the vertical force remains approximately constant as one follows the $\rho_{0,h}-r_{h}$ 
ridge seen in the left panel.

A second, complementary analysis is to fix the various Galactic parameters while 
allowing the shape parameter $(A,B,\theta,\phi,\psi)$ to vary.  This approach is 
very similar to the approach used in many stream studies like \citet{Law2010},
\citet{VeraCiro2013}. and \citet{Koposov2010}.  The PDF for the axis ratios
of this analysis are shown in the lower left panel of Figure 2.
This panel shows that both $A$ and $B$ are poorly 
constrained by the model, with the $1\sigma$ confidence interval on $A$ ranging from 
$1$ to $\simeq 2.5$ and  $B$ ranging from $\simeq 0.4$ to $1$.  Since the
axis ratios are logarithmic, these two ranges are roughly equal.
While some of the inferred models are clearly unphysical, 
they still 
fit the suite of observational constraints that are considered. 
The upper left and lower right panels show that once again, the local vertical 
force remains constant with either parameter.  

\section{Stream Modelling and Data}

In this section we discuss the orbit-fitting technique for 
modelling stellar streams and the production of mock streams of 
an arbitrary angular size.  The mock data are motivated by 
observations of the GD-1 and Orphan streams.

\subsection{The Orbit-Fitting Technique}

The orbit-fitting technique is based on the approximation that stellar streams trace the orbit of a
particle in a fixed potential.  In reality, the kinematics and morphology of the 
stream depends on the evolving tidal field of the host galaxy and the orbit and internal structure 
of the progenitor.  The stream stars are stripped from the progenitor at the inner and 
outer Lagrange points leading to an offset between the orbit and the stream as well as 
a difference between the progenitor's orbital energy and the stream star's energy.
Therefore using the orbit-fitting technique will introduce biases in the inferred quantities
for real streams \citep{Lux2013,Sanders2013a}.
N-body simulations are required
to truly capture the full richness of the stream physics.  However, for an MCMC-based 
parameter estimation scheme, such simulations are prohibitively 
expensive in terms of CPU time.  There are a variety of 
other methods for modelling streams, including, but not limited to, the streak-line 
method \citep{Varghese2011,Kupper2012} and the 
action-angle technique \citep{Sanders2013b}, but they 
introduce additional model parameters and computational challenges.
In this study we generate mock data that lies along the 
orbit and avoid the issue of biases due to the 
use of the orbit-fitting technique.

When using the orbit-fitting technique, one calculates the orbit of a 'progenitor' 
particle in the fixed potential.  If the stream's actual progenitor is 
known the particle is placed at it's location.  Otherwise, it is placed in 
the middle of the observed stream.  The 'progenitor' particle has six-phase space 
coordinates that should be included as free model parameters.  In practice, 
streams often have a thin enough angular thickness that one can fix the 
progenitor's position on the sky with no loss of generality.

Comparing the orbit to the stream data is fairly straightforward.
First the orbital point that has the shortest 
angular distance on the sky to a given data point is found.  This is not a 
trivial step as it requires a search of all calculated orbital points and an 
interpolation between those points that are closest to the data point.
Once the appropriate orbital point is found, either its angular position, 
radial velocity, distance, or proper motions are compared to the 
data point using Eq. 3.2.  

We also model the 'thickness' of the stream in each phase-space coordinate by 
replacing the observed errors with free parameters.  These 'thickness' parameters 
represent the convolution of the observed error with 
the angular stream thickness, the internal velocity dispersion, and the 
spread in distances to the Sun.  

\subsection{Stream Mock Data}

As discussed in the introduction there are about a dozen currently known streams.  The most well-known is the 
Sgr stream.  Its thickness and internal structure make the use of the orbit-fitting 
technique for the stream model problematic.  Instead we use the the GD-1 and Orphan streams.  These 
streams are both long and thin, covering $\sim70^\circ$ and $\sim60^{\circ}$ of the 
sky respectively.  

\citet{Koposov2010} provides observations of the full 6-D phase space for the GD-1 stream.  We will
use those observations as a basis for mock GD-1-like streams.  Similarly, we will use 
the observations of the Orphan's stream angular position, radial velocity, and distance
from \citet{Newberg2010} to generate mock Orphan-like streams.  

The mock streams themselves are made from the orbit of a progenitor particle in the generative 
potential.  The orbit is calculated and the mock data are found by adding the appropriate 
Gaussian error from the real data.  For the angular position, the error is found perpendicularly
to the orbit.  Unlike the other mock data, which is observed at the same location as the 
real data, the mock stream data is spaced evenly along the orbit for an arbitrary angular size.  However, 
the number of data points and type of observations are identical to the real data.  The only difference is 
the length of the stream and the even spacing of the data.

\section{Halo Shape From a Stellar Stream}

We begin by considering the case where the Galactic parameters are known and the 
stream data alone is used to constrain the halo shape parameters.  All other 
parameters are fixed to the generative value, except for the halo scale density,
which is adjust to keep the local circular speed 
equal to the generative model's value of $207~\kms$.
This case, while unrealistic, closely follows the procedure used in 
\citet{Law2010}.

The inferred shape and orientation PDFs for this analysis are shown in the first three 
panels of the left-hand 
column of Figure 3.  
Both the inferred shape and orientation of the halo are highly degenerate.  
These poor constraints are due to the freedom of halo 
orientation.  Since the progenitor location is fixed, the fit to the mock stream 
can be approximated as a fit to the force at the progenitor.  The PDF of this 
force compared to the latitude of the intermediate axis is shown in the 
bottom left-hand panel of Figure 3.  This panel shows that the force at 
GD-1 is almost as tightly constrained as the measurement of the local vertical 
force.  Therefore, as long as the projection of the halo axes at some orientation 
provides roughly the same force at the progenitor location, the model 
will provide an acceptable fit to the stream data.

The second column of Figure 3 shows the results where the constraints from the stream are 
combined with the non-stream constraints (excluding the circular speed).  The improvement 
in the inferred shape constraints is remarkable.  While the constraints on the force at GD-1 have 
decreased by only $\approx 20\%$ compared to the stream alone, the constraints on $A$ and $B$ have
have decreased to $\sim 1-1.8$ and $\sim 0.55-1$, which are roughly equivalent in 
terms of the logarithmic priors.  Evidently, the additional Galactic observations rule 
out many of the other models that fit the GD-1 stream.  

It is worth comparing these results for mock data to the results of \citet{Koposov2010} for 
the actual GD-1 stream.  They were able to find a constraint on the total flattening of the 
potential, but, when the uncertainty in the disk mass was considered, the constraints on the 
halo shape were quite poor.  The right-hand panel of Figure 3 seems to contradict this 
conclusion.  However, we are still fixing all the disk parameters and we are 
using many other observational constraints that were not considered in 
\citet{Koposov2010}.  It is more appropriate to compare the left-hand column of 
Figure 3 to their results.  While a direct comparison is difficult, both our results and 
their Figure 18 show very poor constraints on the halo shape.  It seems that the 
uncertainty introduced by the halo orientation affects the inferred shape in a similar manner 
as the uncertainty in the disk mass.  However, it must be remembered that they are analyzing 
a stream while we are analyzing mock data generated from an orbit, which may also account for 
some of the differences in the quality of the inferences.

The PDFs for the GD-1-like stream plus the Galactic constraints represent a 
best-case scenario for stream modelling.  The stream is traced by the orbit of a 
single particle and the majority of the Galactic parameters are known.  Yet both 
$A$ and $B$ are still uncertain at the $\pm50\%$ level.  The stream has significantly 
improved the constraints on the halo shape, but, when the halo orientation is 
allowed to vary, the uncertainties in the shape parameters are still quite large.  

In reality the precise values of the various MW model parameters are not known
and it is certainly possible that the use of incorrect values for 
the fixed model parameters may introduce systematic errors in the 
parameters of interest.  To illustrate this point, let us consider a common assumption in 
studies of stellar streams: that the disk and halo are aligned.  To that 
end, we re-analyze our mock data, which were constructed from a 
model where the
minor axis of the halo is the closest to the disk's spin axis but differs
from the spin axis by $30^\circ$, under this assumption.  
In this re-analysis we allow all the parameters to vary except two of the Euler angles, forcing 
one of the halo symmetry axes to be parallel to the disk spin axis.
We fit all the observational constraints, including the circular speed and the 
GD-1-like stream.
For simplicity, we replace $A$ and $B$ with $A^{'}$ and $B^{'}$ where $A^{'}$ is the 
ratio of the larger axis in the plane of the disk to the smaller axis in the plane of the 
disk and $B^{'}$ is the ratio of the axis parallel to the disk spin axis to the smallest
axis in the plane of the disk.  The value of the smallest axis in the plane of the disk is 
set to one.  Therefore, $A^{'}$ can refer to either the major or intermediate axis and 
$B^{'}$ can be the major, minor, or intermediate axis.

The PDF for $A^{'}$ and $B^{'}$ is shown in Figure 4.  Any model with $B^{'}<1$ has the
minor axis parallel to the disk spin axis while any model with $B^{'}>A^{'}$ has the major 
axis parallel to the disk spin axis.  Figure 4 shows that our analysis overwhelmingly prefers
models with the intermediate axis parallel to the disk spin axis.  This disagrees with 
the generative model, which has the minor axis closest to the disk spin axis.

The result from this analysis may resolve a discrepancy in the literature about the Sgr stream.
Both \citet{Law2010} and \citet{DW13} considered triaxial halos and assumed that the 
disk and halo were aligned.  Using the Sgr stream, both studies found the 
intermediate axis of the halo was parallel to the disk spin axis.  However,
simulations by \citet{Debattista2013} show that such an alignment is 
unstable.  Our results suggest that 
the alignment found by \citet{Law2010} and \citet{DW13} may be due to the assumption 
of that the disk and halo are aligned.  We also note that this discrepancy may be 
caused by other assumptions in \citet{Law2010} and \citet{DW13}.  For instance, 
\citet{VeraCiro2013} showed fit the Sgr stream using a halo where the triaxiality 
varies with radius.  Alternatively, \citet{Ibata2013} used a non-parametric spherical 
halo to fit the Sgr stream.  Neither of these studies have results that disagree with 
\citet{Debattista2013}.  Furthermore, no study of the Sgr stream has been able to 
fit all the known Sgr data, including the bifurcation.

In order to avoid confusion and systematics that may arise due to incorrect assumptions 
about the model parameters, one should allow all the model parameters to vary according 
to their priors while fitting all possible observational constraints simultaneously.
Figure 5 shows the results for an analysis where we follow such a procedure.  
These PDFs are marginally smaller than the for the suite of constraints alone and 
significantly smaller than for the GD-1 stream alone.  In both of those cases the 
Galactic parameters were fixed, while in this case the parameters are completely free.
When one considers the increased freedom, the improvement, whether marginal or 
significant, over an analysis with the Galactic parameters fixed reinforces 
the point of Figure 3.  Constraints from stellar streams should be incorporated into 
a larger suite of Galactic constraints when modelling the MW.

As expected, the PDFs for the shape and orientation have increased in size relative 
to the second column of Figure 3.  The $1~\sigma$ limits on $A$ and $B$ are now 
$\sim 1-2$ and $\sim 0.4-1$ compared to Figure 3's limits $\sim 1-1.8$ and $\sim 0.55-1$.
The area of the $1~\sigma$ shape contours in second column of Figure 3 is 
$\approx 60\%$ of those in Figure 5.
Nonetheless the contours of Figure 5 represent what can realistically be achieved 
in a best case scenario with a GD-1-like stream using the orbit-fitting method.  

\section{Long Streams and Multiple Streams}

Further observations may provide avenues to break the degeneracies in the 
shape parameters found in Figure 5.
Firstly, one could use a single longer stellar stream in the model constraints.  
The longer stream should improve the constraints by reducing the number 
of possible orbits that fit the data as many of the orbits that fit a
shorter stream diverge rapidly past the edge of the data.

To explore how much 
can be gained with an extended stream we generated a $180^\circ$ GD-1-like stream.  Combining that 
stream with the suite of other constraints we repeat the analysis of Section 5.  The resulting 
shape PDFs are shown in Figure 6.  In this case the $1~\sigma$ limits on $A$ and $B$ have been 
reduced to $\simeq 1-1.5$ and $\simeq 0.65-1$ respectively.  This gives an area for the shape contours 
that is $\approx 35\%$ of the area seen in Figure 5.  The shape contours for 
the $180^\circ$ stream where the full parameter space is explored are roughly 
equal to the inference with a $70^\circ$ where only the shape and orientation parameters 
are explored.

A second idea to remove some of the degeneracies present in the model parameters is simply to utilize 
multiple stellar streams simultaneously.  As a proof-of-concept, let us consider the
$70^\circ$ GD-1-like stream used previously and a $60^\circ$ Orphan-like stream.
Before combining the constraints from the two streams, we analyzed the Orphan-like 
stream together with the Galactic constraints.  As with the GD-1 stream, we fix the angular 
position of the progenitor particle to the middle of the mock stream while varying the 
other four phase-space coordinates.
The shape and orientation PDFs for that 
analysis are shown in Figure 7.  In this case the constraints 
on the shape and orientation are worse than those for GD-1.  However, it should be noted that the mock data for the Orphan-like 
stream contains significantly fewer data points that the GD-1-like stream and does 
not include proper motions. 

The PDFs for the combination of the two streams are shown in Figure 8.  
The shape contours have improved from those of Figure 4, while the 
orientation contours  appear to be slightly worse.  Here the absolute $1~\sigma$ limits on 
$A$ and $B$ are $\simeq 1- 1.6$ and $\simeq 0.5-1$ respectively.  
Additionally, the PDF contains a ridge that reduces the area of the $1~\sigma$ contours.
Ultimately, area contained in axis ratio 
contours for two streams is equivalent to the area for a single $180^\circ$ long GD-1-like 
stream.

\section{Conclusions}

In this work we have examined how stellar streams can be used to remove some of the degeneracies 
found in MW models.  In our view one of the main takeaway points of this study is that 
in modelling the MW, one should apply Galactic and stream constraints simultaneously
while exploring the full parameter space.  Figure 3 clearly shows that 
the full suite of Galactic observations, which include 
local measurements of the vertical force, circular speed, and Oort 
constants provide useful information on the halo shape.  
A comparison of Figure 5 and Figure 3 shows that the 
standard practice whereby one fixes the structural 
parameters of the disk, bulge, and spherically-averaged halo while fitting 
the axis ratios and orientation of the halo may lead one to over-estimate 
the extent to which stream observations constrain the shape of the dark 
halo. 

A second lesson is one should avoid restrictive model assumptions 
such as an alignment between the symmetry axis of the disk and one of the 
symmetry axes of the halo.  Indeed, it appears that it is precisely this 
assumption that lead \citet{Law2010} and \citet{DW13} to the 
unphysically unstable model wherein the disk sits in the plane defined by 
the major and minor axis \citep{Debattista2013}.  When we relax the assumption 
of disk-halo alignment and fully explore the parameter space we
are able to recover the shape and orientation of the 
halo with uncertainties on the order of $\simeq50\%$.  In this 
respect, our paper complements that of 
\citet{Ibata2013}, where the lesson is that models that assume a particular form 
for the spherically-averaged halo density profile may be too restrictive.

The third point of this paper is that the use of extended streams or 
multiple streams improve the constraints on the halo shape significantly.
However, our analysis of mock data suggests that, at best, combining 
observation of the GD-1 stream with the Orphan stream and other Galactic 
observations, cannot constrain the halo shape to better than
$\simeq25\%$.

A large portion of this uncertainty arises from 
our consideration of the halo orientation.  A single, short stream is mostly 
sensitive to the projection of the halo onto the stream.  This lack of sensitivity to 
the total halo gives a great deal of freedom to its axis ratios.  As the stream size 
increases, or as more streams are considered, a greater portion of the Galaxy is 
explored, giving more sensitivity to the total halo and 
reducing the uncertainty in the halo parameters. Related to this point, recent studies 
suggest that streams containing turning points are particularly sensitive 
the halo shape \citep{Varghese2011,Lux2012}.  In this work we only considered mock GD-1 and 
Orphan streams which do not contain such turning points.  Other, shorter, streams like 
Pal 5 and NGC 5466 may be more promising than our results suggest if turning points in the 
streams are found.

These results suggest that the path to improved inferences of the halo structure 
is to fit all available streams and other Galactic observations 
simultaneously.  There remains, however, the
issue that the map from gravitational potential to stream is not
as simply as the orbit-fitting technique deployed in this paper.
How one explores the full range of Galactic models while properly 
accounting for stream dynamics remains an outstanding challenge.

Nonetheless, the central points of this work remains true regardless of the stream modelling 
technique.  Stellar streams provide powerful probes of the Galactic structure, but they 
must be combined with a suite of other observational constraints to reach their full 
potential.  It is necessary to fully explore the model parameter space as model 
degeneracies may cause errors in the inferred quantities depending on the assumed 
quantities.  Finally, the degeneracies in the model parameters can be reduced 
through the use of extended streams or multiple streams, yielding better models of 
the MW.  

L.M.W. is supported by the Natural Sciences and Engineering
 Research Council of Canada through a Discovery Grant.

\section{Appendix: EMCEE on a Sphere}

Our Galactic model has three Euler angles and triaxial symmetry.  The first two 
Euler angles specify the angular position of the intermediate axis.  Therefore, when 
the EMCEE algorithm stretches between two walkers it should be along the great circle 
connecting the two intermediate axes.  Additionally, the triaxial symmetry introduces 
a number of degeneracies in the Euler angles.  Since the pole and anti-pole of an axis 
provide identical models, the stretch should be along the shortest arc connecting the 
pole or anti-pole of the second walker's intermediate axis to the first walker
axis.

To achieve this, we modify the EMCEE algorithm in a straightforward manner.  For a 
given pair of walkers, the location of the intermediate axes on a sphere are found.
These two points are then rotated to a coordinate system,  $(\lambda^{'},\beta^{'})$,
that has one of the points at $\beta^{'}=90^{\circ}$.  In this coordinate system, the 
great circle connecting the two points has a constant $\lambda^{'}$ so  the stretch 
is only in $\beta^{'}$.  If the second walker has $\beta^{'}> 0$ then the stretch is 
between those two points.  Otherwise, we replace the second point with its anti-pole for 
the stretch.  The proposal is then generated along the arc connecting the appropriate 
pair of points and rotated back to the original coordinate system to give the 
proposal Euler angles.  This technique removes the degeneracies in the first two Euler 
angles and stretches along the great circle between two points on a sphere.

\begin{figure*}
\centering
\begin{minipage}{126mm}
\includegraphics[width=126mm]{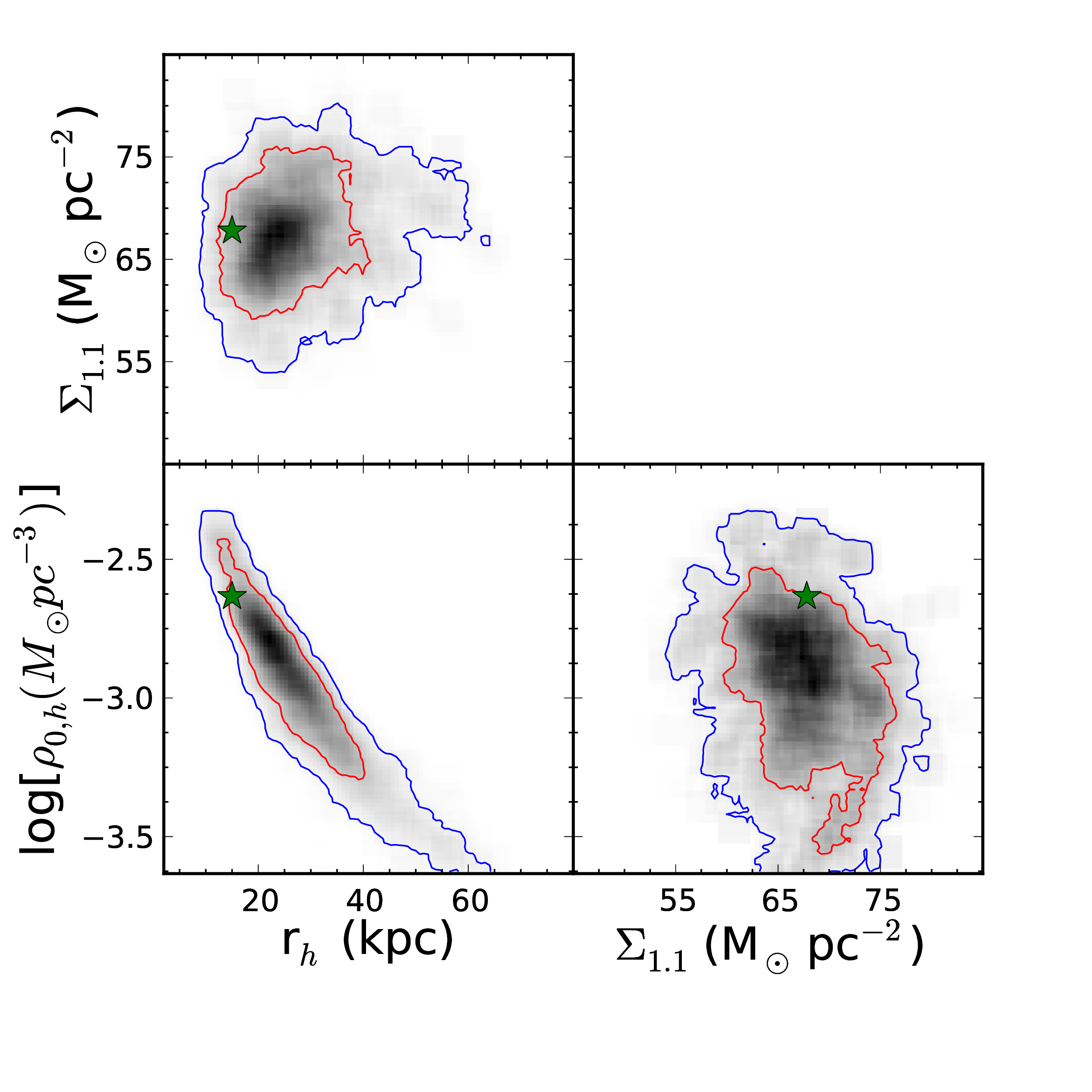}
\caption{PDFs of the halo scale density, halo scale length, and local vertical force, $\Sigma_{1.1}$ where
$\Sigma_{1.1}=|K_{z}(1.1\textrm{kpc})|/2\pi G$.
The lower left panel shows the 
relation between the halo scale density and scale length, the upper left panel shows the local vertical force 
and the scale length, and the lower right panel shows the local vertical force and scale density.  The red and blue
contour lines are the one and two sigma limits of the PDFs. The green star indicates the value of the generative 
parameter and the data constraint.}
\end{minipage}
\end{figure*}

\begin{figure*}
\centering
\begin{minipage}{126mm}
\includegraphics[width=126mm]{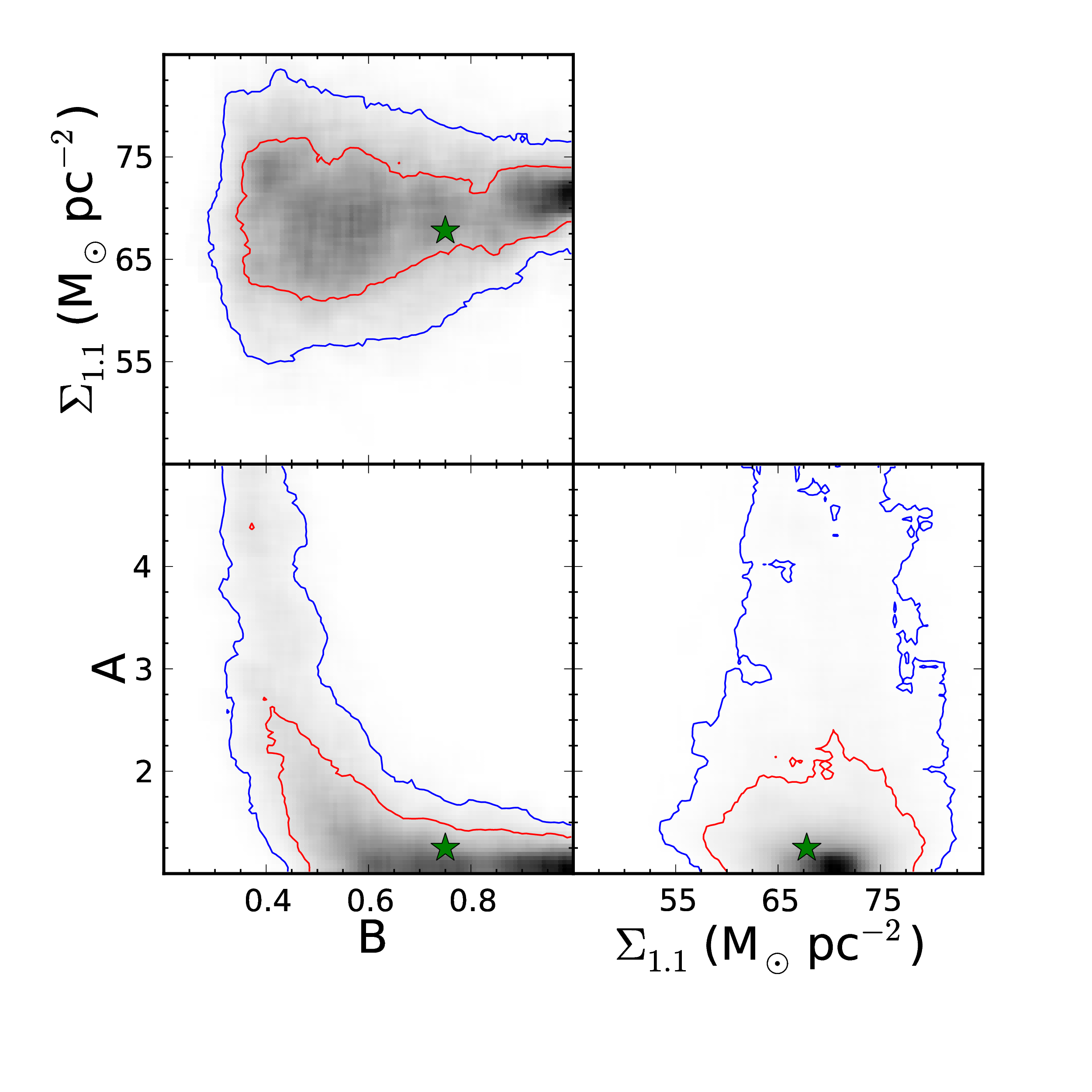}
\caption{PDFs of the halo axis ratios and local vertical force. The lower left panel shows the 
relation between the $A$ and $B$, the upper left panel shows the local vertical force 
and $B$, and the lower right panel shows the local vertical force and $A$.  The contours,
green stars, and $\Sigma_{1.1}$ are as in Figure 1.}
\end{minipage}
\end{figure*}

\begin{figure*}
\centering
\begin{minipage}{126mm}
\includegraphics[width=126mm]{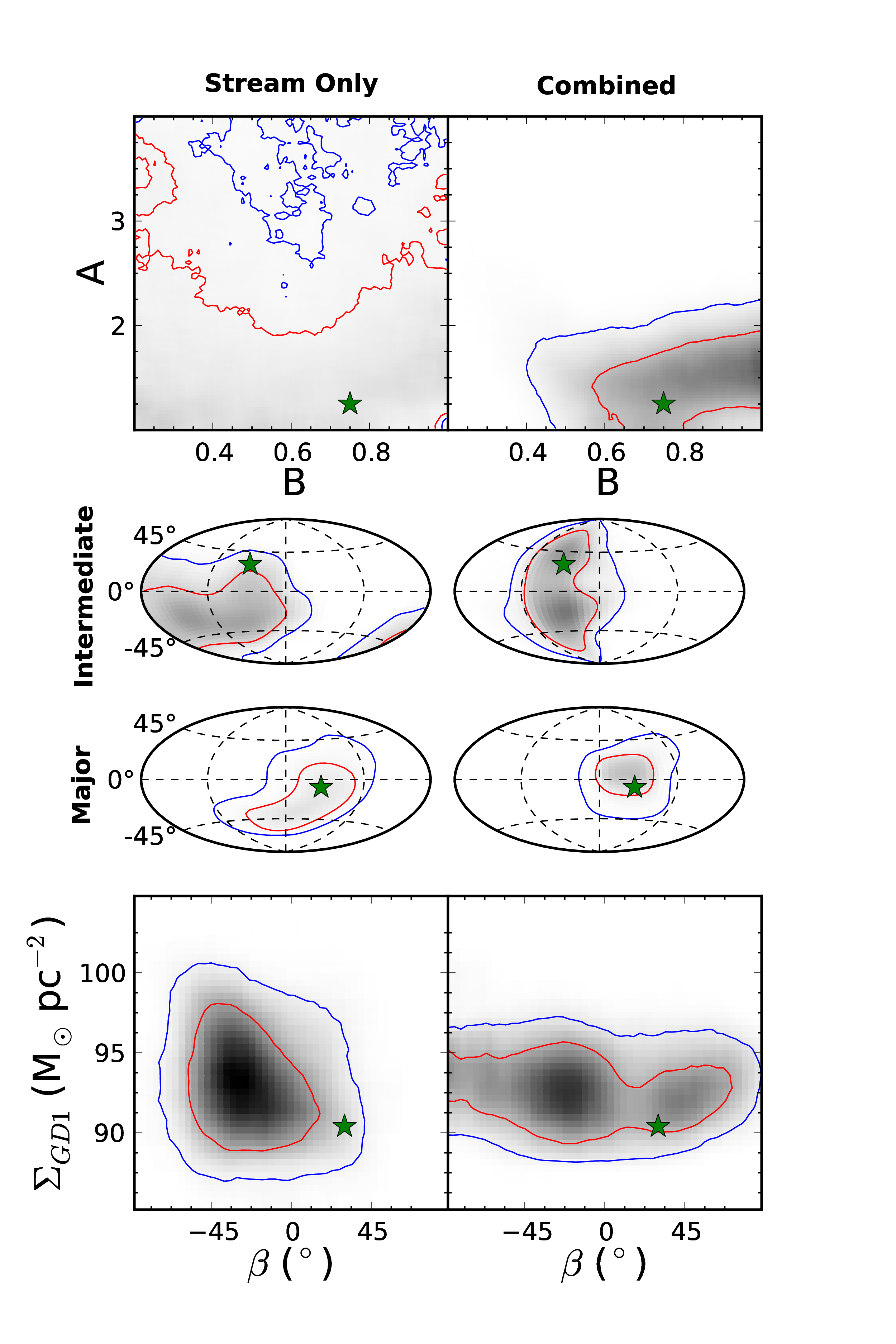}
\caption{The inferred shape, orientation and force at the GD-1 progenitor particle's location PDFs. 
In this figure $\Sigma_{GD1}=|F|(\textrm{GD-1})/2\pi G$.
The first column uses only the stream information. 
The second column uses both the stream and non-stream data 
simultaneously.  The first row shows the inferred shape.  $A$ is the major/intermediate axis ratio and
$B$ is the minor/intermediate ratio.  The second row shows the location of the intermediate axis in angular
coordinates centered on the Galactic Center with the Sun at along the $(0^\circ,0^{\circ})$ line.  The 
third row shows the location of the major axis in that same coordinate system.  The minor axis lies perpendicular
to the major and intermediate axes.  The bottom row shows the total force at the GD-1 progenitor 
compared to the latitude of the inferred intermediate axis.  The units of the 
force are $\msol~\textrm{pc}^{-2}$.
The green and contours are as in Figure 1.  Each row uses the same color map limits to highlight 
the differences between the use of the different sets of constraints.  Additionally the bottom three rows 
have had the pole and anti-pole points stacked together as they are equivalent.  The dashed black lines in the 
Aitoff plots are at longitudes of ($-90^\circ,0^\circ,90^\circ$) and latitudes of ($-45^\circ,0^\circ,45^\circ$). }
\end{minipage}
\end{figure*}

\begin{figure*}
\centering
\begin{minipage}{126mm}
\includegraphics[width=126mm]{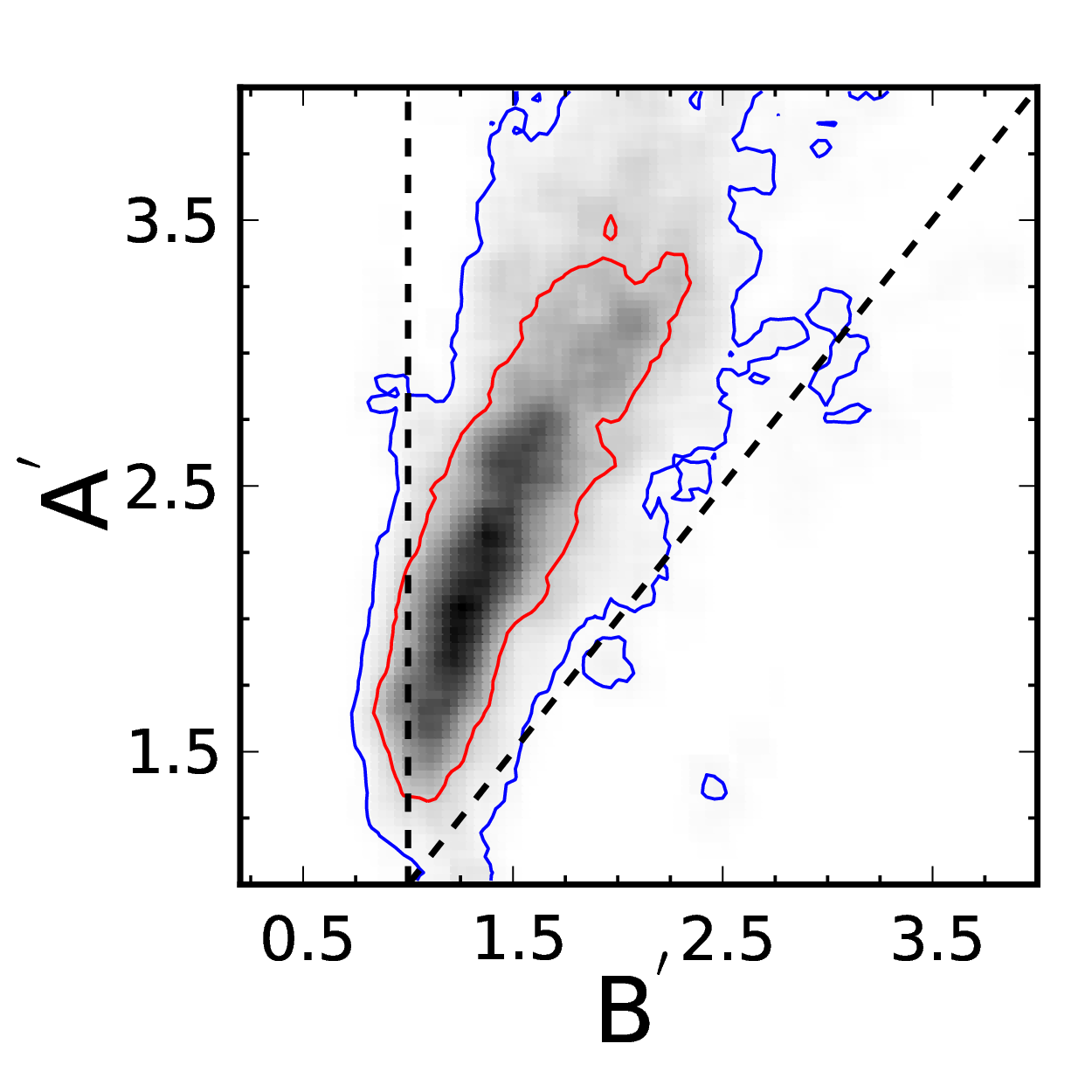}
\caption{The inferred shape using the $70^\circ$ GD-1 stream and the standard suite of constraints.  In this analysis 
all parameters are allowed to vary with the exception that one axis lies perpendicular to the disk and the other 
two are in the plane of the disk.  $A^{'}$ is ratio of the larger to the smaller disk axes.  $B^{'}$ is the 
ratio of the perpendicular axis to the smaller disk axis.  The contours are the same as in Figure 1.
The dashed lines separate the parameter space into regions where the perpendicular axis is the minor axis ($B^{'}<1$)
and where the perpendicular axis is the major axis ($B^{'}>A^{'}$).}
\end{minipage}
\end{figure*}

\begin{figure*}
\centering
\begin{minipage}{126mm}
\includegraphics[width=126mm]{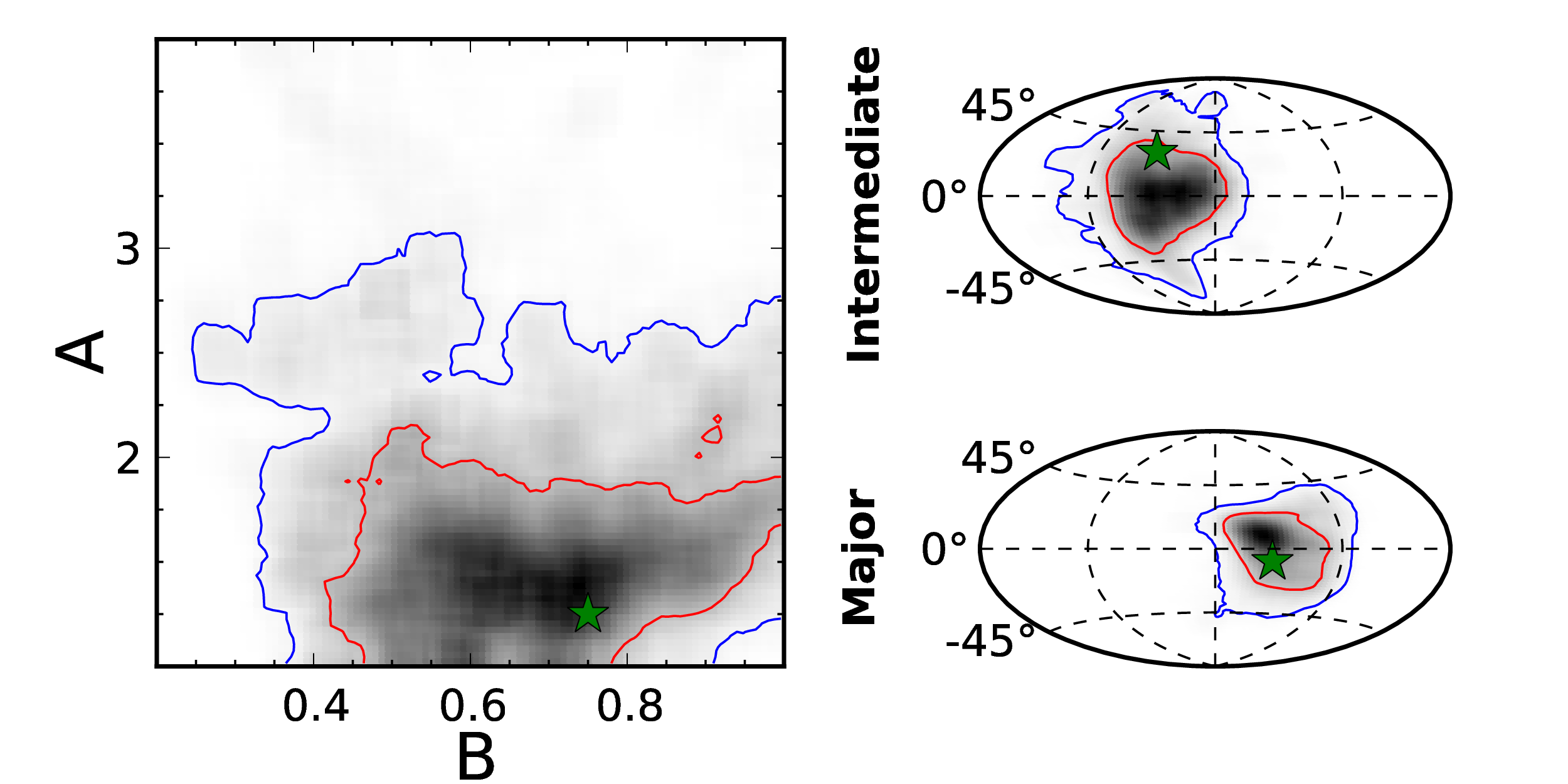}
\caption{The inferred shape and orientation PDFs using the $70^\circ$ GD-1 stream and the standard suite of 
constraints and allowing all parameters to vary.  The contours and green stars are as in Figure 1 and the Aitoff axes are as in Figure 3.}
\end{minipage}
\end{figure*}

\begin{figure*}
\centering
\begin{minipage}{126mm}
\includegraphics[width=126mm]{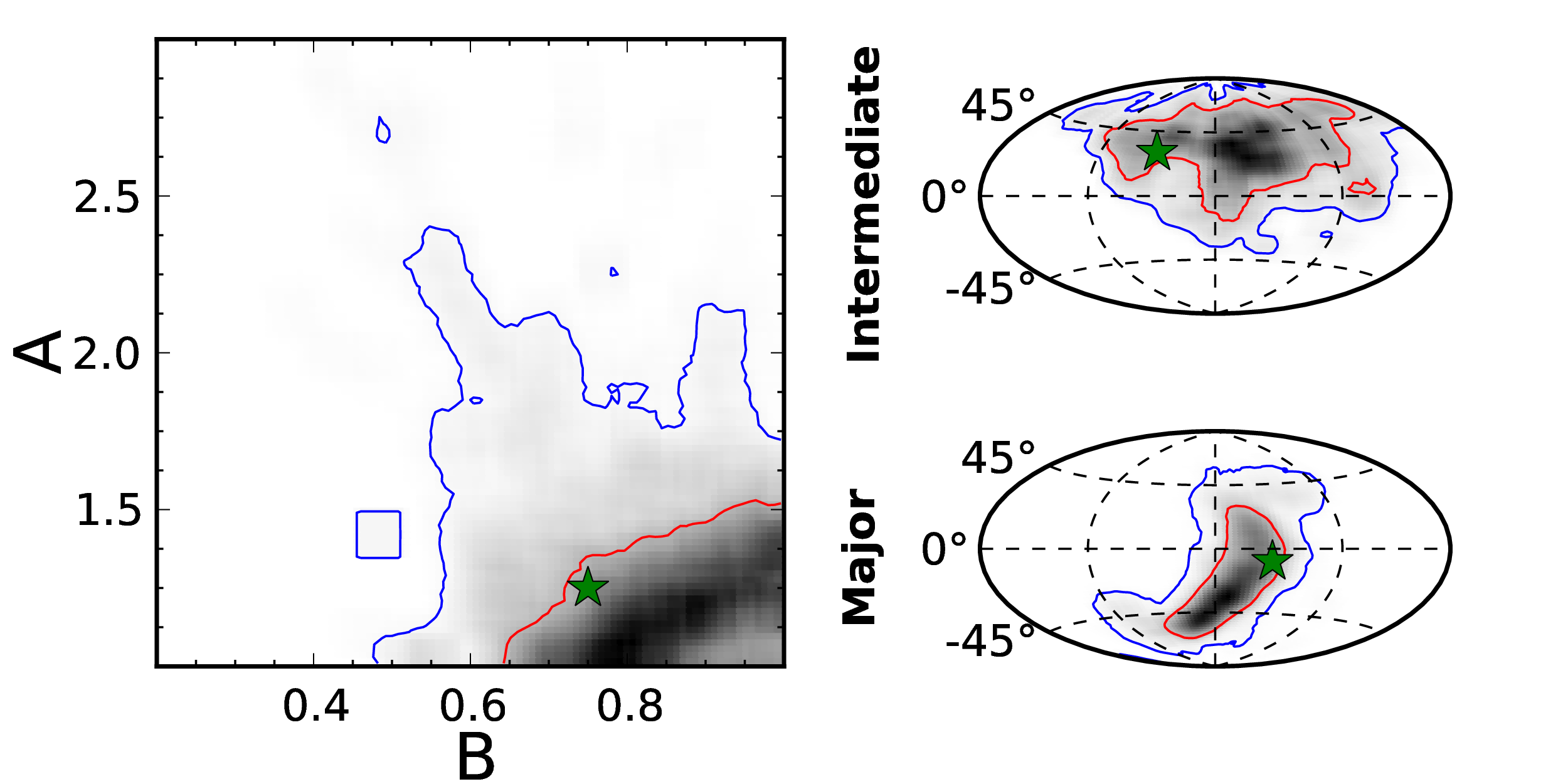}
\caption{The inferred shape and orientation PDFs using the $180^\circ$ GD-1 stream and the standard suite of 
constraints and allowing all parameters to vary.  The contours and green stars are as in Figure 1 and the Aitoff axes are as in Figure 3.}
\end{minipage}
\end{figure*}

\begin{figure*}
\centering
\begin{minipage}{126mm}
\includegraphics[width=126mm]{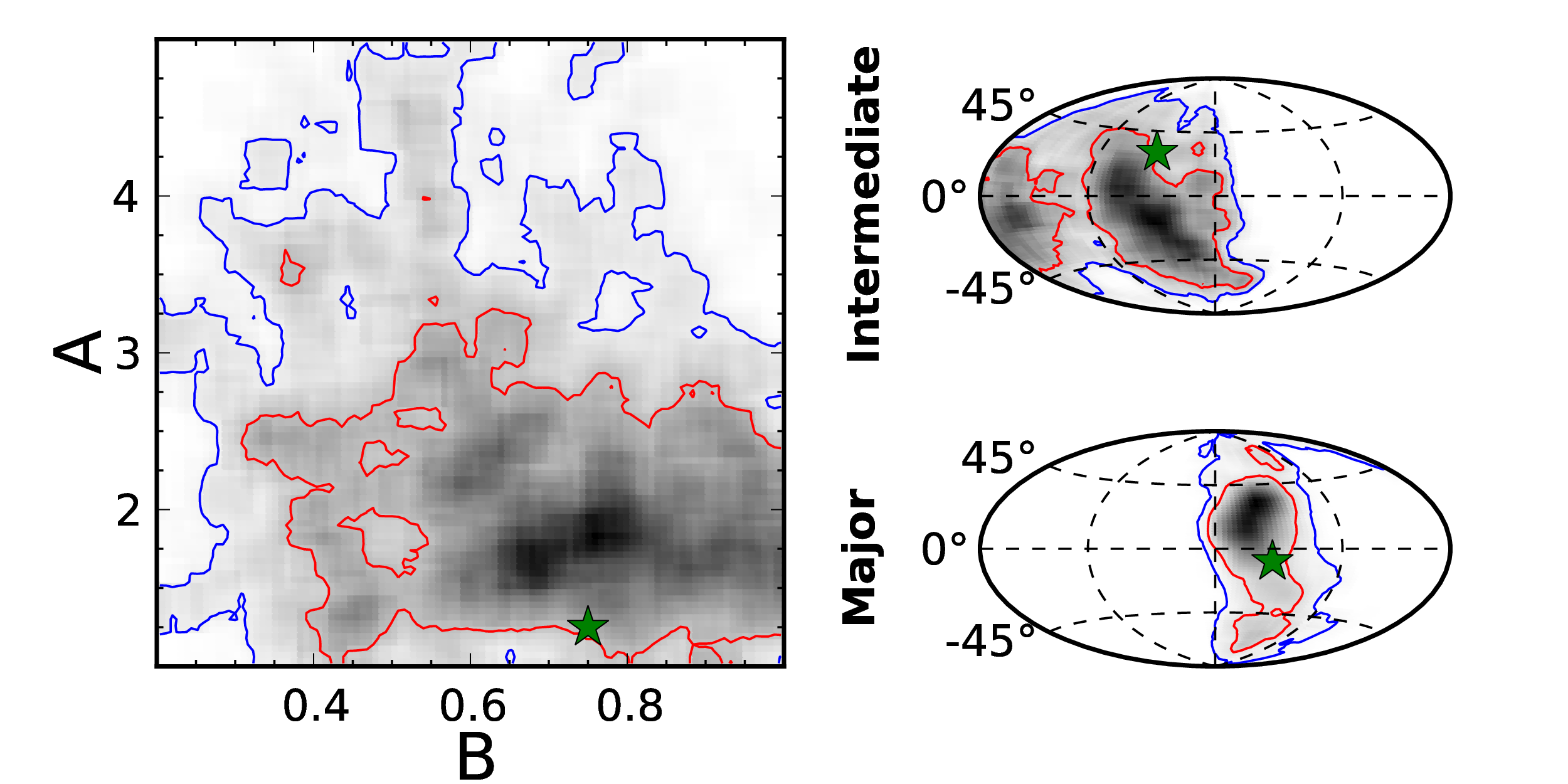}
\caption{The inferred shape and orientation PDFs using the $60^\circ$ Orphan-like stream and the standard suite of 
constraints and allowing all parameters to vary.  The contours and green stars are as in Figure 1 and the Aitoff axes are as in Figure 3.}
\end{minipage}
\end{figure*}

\begin{figure*}
\centering
\begin{minipage}{126mm}
\includegraphics[width=126mm]{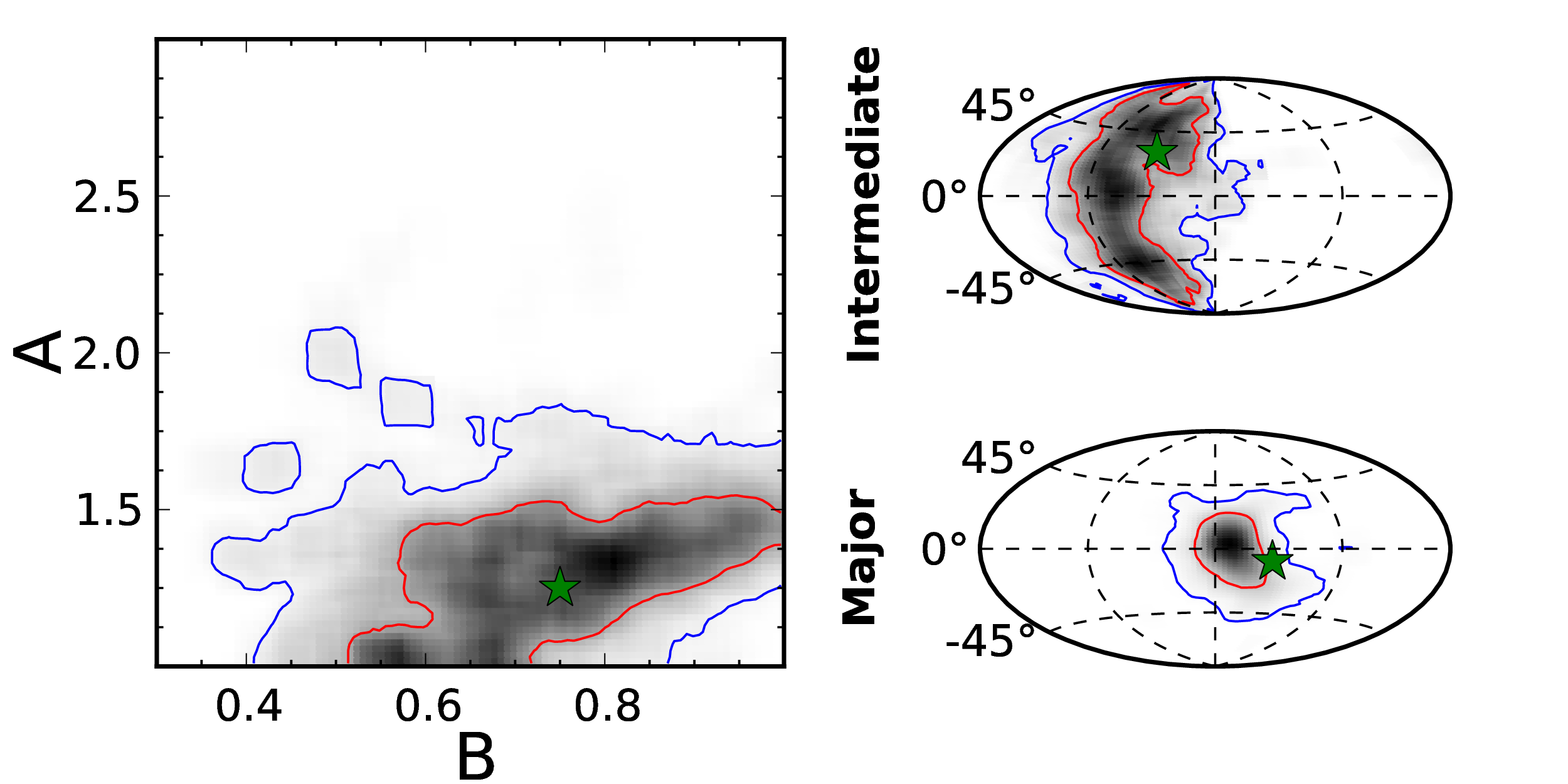}
\caption{The inferred shape and orientation PDFs using the $70^\circ$ GD-1 and the
$60^\circ$ Orphan streams and the standard suite of 
constraints and allowing all parameters to vary.  The contours and green stars are as in Figure 1 and the Aitoff axes are as in Figure 3.}
\end{minipage}
\end{figure*}

\end{document}